# A New Dynamic Round Robin and SRTN Algorithm with Variable Original Time Slice and Intelligent Time Slice for Soft Real Time Systems

### H.S. Behera
Sr. Lecturer
Veer Surendra Sai University Of
Technology, Burla
Sambalpur, India

### Simpi Patel
Student
Veer Surendra Sai University Of
Technology, Burla
Sambalpur, India

### Bijayalakshmi Panda
Student
Veer Surendra Sai University Of
Technology, Burla
Sambalpur, India

## ABSTRACT

The main objective of the paper is to improve the Round Robin (RR) algorithm using dynamic ITS by coalescing it with Shortest Remaining Time Next (SRTN) algorithm thus reducing the average waiting time, average turnaround time and the number of context switches. The original time slice has been calculated for each process based on its burst time.This is mostly suited for soft real time systems where meeting of deadlines is desirable to increase its performance. The advantage is that processes that are closer to their remaining completion time will get more chances to execute and leave the ready queue. This will reduce the number of processes in the ready queue by knocking out short jobs relatively faster in a hope to reduce the average waiting time, turn around time and number of context switches. This paper improves the algorithm [8] and the experimental analysis shows that the proposed algorithm performs better than algorithm [6] and [8] when the processes are having an increasing order, decreasing order and random order of burst time.

## 1. INTRODUCTION

Real time systems are systems which react to the events in the surrounding by carrying out specific actions within the specified time. A real-time deadline can be so small that system reaction appears instantaneous. The term real-time computing has also been used however, to describe "slow real-time" output that has a longer, but fixed, time limit. There are three types of real-time systems. The types of real-time systems include hard, soft and adaptive real-time systems. Hard real time system says that all of the deadlines or temporal constraints have to be resolved. Second type of this system, which is known as soft real-time system suggests that missing single deadline should not put the system behaviour in danger. It often denotes a system that attempts to meet all time constraints imposed by its tasks or operations or applications by enjoying the powerful system resources such as high clock rate, faster processors, speedy cache, and lightening buses. It is still a 'soft' real-time system because some critical tasks might be delayed due to some system-oriented processes that are bulky and time-consuming and not preemptive.Adaptive real-time system adjusts the internal strategies by giving response to the changes that are carried out in the environment.

## 1.1 Preliminaries

A program in execution is called a *process*. The processes, waiting to be assigned to a processor are put in a queue called ready queue. CPU Utilization is the capacity to keep the CPU busy as much as possible as long as there are jobs to process. Throughput is a measure of work in terms of the number of processes that are completed per unit time for which a process holds the CPU is known as burst time. The time at which a process arrives is its arrival time. *Turnaround time* is the amount of time to execute a particular process, while *waiting time* is the amount of time a process has been waiting in the ready queue. Time elapsed between the submissions of a request by the process till its first response is called the response time. In time sharing system, the CPU executes multiple processes by switching among them very fast. The number of times CPU switches from one process to another is called the number of *context switches*.

Scheduling disciplines are algorithms used for distributing resources among parties which simultaneously and asynchronously request them. Scheduling disciplines are used in routers (to handle packet traffic) as well as in operating systems (to share CPU time among both threads and processes), disk drives (I/O scheduling), printers (print spooler), most embedded systems, etc. The main purposes of scheduling algorithms is to minimize resource starvation, to ensure fairness amongst the parties utilizing the resources and to keep the CPU busy as much as possible by executing a (user) process and then switching to another process . Scheduling deals with the problem of deciding which of the outstanding requests is to be allocated resources.

In general, (job) scheduling is performed in three stages: short-, medium-, and long-term. The activity frequency of these stages is implied by their names.Long-term (job) scheduling is done when a new process is created. It initiates processes and so controls the degree of multi-programming (number of processes in memory). Medium-term scheduling involves suspending or resuming processes by swapping (rolling) them out of or into memory. Short-term (process or CPU) scheduling occurs most frequently and decides which process to execute next.

## 1.2 Scheduling Policies

In general, scheduling policies may be *preemptive* or *non-preemptive*. In a non-preemptive pure multiprogramming system, the short-term scheduler lets the current process run until it blocks, waiting for an event or a resource, or it terminates i.e First-Come-First-Served (FCFS), Shortest Job first (SJF) policies.

### 1.2.1 First come first served
It is the simplest scheduling algorithm, FIFO simply queues processes in the order that they arrive in the ready queue.

### 1.2.2 Shortest job first
With this strategy the scheduler arranges processes with the least estimated processing time remaining to be next in the queue. This requires advance knowledge or estimations about the time required for a process to complete.

In a *preemptive* multiprogramming system, the short term schedular permits a process to be removed from processor when other high priority process enters into the system.

### 1.2.3 Fixed priority pre-emptive scheduling
It is a scheduling system commonly used in real-time systems. With fixed priority pre-emptive scheduling, the scheduler ensures that at any given time, the processor executes the highest priority task of all those tasks that are currently ready to execute.

### 1.2.4 Rate-monotonic scheduling
A scheduling algorithm used in real-time operating systems with a static-priority scheduling class. The static priorities are assigned on the basis of the cycle duration of the job: the shorter the cycle duration is, the higher is the job's priority.

### 1.2.5 Round-robin scheduling
Round-robin (RR) is one of the simplest scheduling algorithms for processes in an operating system which assigns time slices to each process in equal portions and in circular order,handling all the processes without priority (also known as cyclic executive). Round-robin scheduling is both simple and easy to implement, and starvation free.

### 1.2.6 Earliest deadline first
Earliest Deadline First (EDF) or Least Time to Go is a dynamic scheduling algorithm used in real-time operating systems. It places processes in a priority queue. Whenever a scheduling event occurs (task finishes, new task released, etc.) the queue will be searched for the process closest to its deadline. This process is the next to be scheduled for execution.

## 1.3 Related work

[5]and[6] Yaashuwanth.C & R.Ramesh proposed an architecture which eliminates the defects of implementing a simple round robin architecture in real time operating system by introducing a concept called intelligent time slicing which depends on three aspects i.e. priority, average CPU burst and context switch avoidance time.[7] Prof. Rakesh Mohanty, Prof. H. S. Behera et.al proposed a new Improved-RR algorithm named Shortest Remaining Burst Round Robin Scheduling Algorithm(SRBRR). [8] Prof. Rakesh Mohanty, Prof. H. S. Behera et.al proposed Priority Based Dynamic

Round Robin (PBDRR) Algorithm with Intelligent Time Slice for Soft Real Time Systems.

## 1.4 Our Contribution

The original time slice suited to the burst time of each process as mentioned in [5] has been calculated, The dynamic ITS as in [6] and [8] has been found out and RR in conjunction with the SRTN algorithm both of which are pre-emptive in nature (suitable for soft real time system) have been used and it is observed that there is a further improvement in the performance metrics

## 1.5 Organization of the paper

Section 2 presents the pseudo code and illustration of our proposed algorithm .In section 3 experimental analysis of the proposed algorithm and its comparison with the algorithms in [6] and [8] is presented. Section 4 contains the conclusion and future work.

## 2. PROPOSED ALGORITHM
## 2.1 Uniqueness of the Approach

In the proposed algorithm jobs are assigned original time slice based on the burst time of each process and intelligent time slice for each cycle. RR with SRTN has been used because performance of RR soley depends on time quantum.If it is too small it causes context switches. If it is very large it degenerates the algorithm to that of FCFS. Because SRTN allows the execution of jobs with shortest remaining time first, hence it allows shorter processes to leave the queue thus allowing faster execution of fewer processes.

## 2.2 Detailed Structure of the algorithm

First, the original time slice (OTS) to be allocated to each process is calculated by using a formula which takes in to account the range, priority and total number of processes in the CPU.OTS has been assigned based on the priority of each process. Then the ITS based on OTS, priority component, shortest CPU burst time and context switch component has been found out. RR along with SRTN has been used for scheduling the processes. Performance of RR depends on the time quantum while that of SRTN depends on quicker execution of the processes with least remaining burst time. If time quantum is too small it results in context switch overhead resulting in loss of precious CPU time while a large time quantum degenerates it to FCFS algorithm.

### 2.2.1 Pseudo code:
1. Let n : number of processes
   bt[i] : burst time of ith process.
   rbt[i] : remaining burst time of ith process
   r : number of the round
   Initialize: cs=0, $avg_{wt}$=0, $avg_{tat}$=0,r=1
2. Calculate OTS for all n processes present in the ready Queue
   //OTS is the original time slice//
       Range = maximum CPU burst time + minimum CPU burst time
   //Range is the range of burst time of n processes//
       OTS = (Range*Total number of processes in the System)/ (Priority of the process* Total Number of priorities in the system)
   //Priority is the user defined priority of the processes//
2. Calculate the ITS for all n processes in the ready queue

//ITS is the intelligent time slice of the processes//

3. Arrange all the processes in the ready queue in ascending order of rbt[i]

4. While (ready queue! = NULL)

if ( r==1)

$$TQ = \begin{cases} \frac{1}{2} ITS, \text{ if } SC= 0 \\ ITS, \text{ otherwise} \end{cases}$$

else

$$TQ = \begin{cases} TQ_{i-1} + \frac{1}{2} TQ_{i-1}, \text{ if } SC=0 \\ 2 * TQ_{i-1}, \text{ otherwise} \end{cases}$$

if (rbt[i] - TQ_i) <=2

$TQ_i= bt[i]$

//TQ is the time quantum assigned to each process//

5. Assign TQ to as the burst time of process i

$bt[i] \rightarrow TQ_i$

6. If (i < n)

i=i+1

goto step (4)

else

update counter r and goto step (3)

End of while

7. cs, $avg_{wt}$, $avg_{tat}$ are calculated.

8. End

### .2.3 Illustration

Given the burst sequence: 25 60 12 43 5 with user priority 3 1 2 1 1 respectively. Range was found out by adding the highest CPU burst time and the smallest CPU burst time and dividing the result by 2. Original Time Slice (OTS) was then calculated by dividing the range of processes multiplied to total number of processes and priority of each process multiplied to total number of priorities. It was found to be 11 33 17 33 33. The priority component (PC) is assigned 0 or 1 depending upon the priority assigned by the user which is inversely proportional to the priority number .It was calculated as 0 1 0 1 1. Shortness component (SC) difference between burst time of current process and its previous process is calculated .It is 1 if difference is less than zero,0 otherwise. (SC) was calculated and was found to be 0 0 1 0 1. If this balance CPU burst is less than OTS, it will be considered as Context Switch Component (CSC) otherwise it isn't considered as CSC. The CSC was calculated as 0 0 12 9 5. The intelligent time slice is sumof all the values like OTS, PC, SC and CSC. Intelligent time slice for individual processes was computed as 11 34 30 43 5.The processes are then arranged in increasing order of their burst time. In first round, the processes having SC as 1 were assigned time quantum same as intelligent time slice whereas the processes having SC as 0 were given the time quantum equal to the roof of the half of the intelligent time slice. So processes P1, P2, P3, P4, P5 were assigned time quantum as 6 17 12 22   5.The remaining burst times were found out and the processes were again arranged in increasing order of their burst time following the SRTN scheduling .  In next round , processes having SC as 1 were assigned double slice of its previous round whereas the Processes with SC equals to 0  were given the time quantum equal to  the sum of the Previous time quantum and roof of the half of the previous time quantum. Similarly time quantum is assigned to each process available in each round of execution.

## 3. EXPERIMENTAL ANALYSIS

### 3.1 Assumptions

All the experiments are performed is a single processor environment and all the processes are independent. Attributes like burst time, priority, number of processes is known before submitting the processes to the processor. All processes are CPU bound. No processes are I/O bound.

### 3.2 Experimental Framework

The experiment consists of several input and output parameters. The input parameters consist of burst time, priority and the number of processes. The output parameters consist of average waiting time, average turnaround time and number of context switches.

### 3.3 Data Set

Several experiments have been performed for evaluating performance of the new proposed algorithm but only three of them are shown .The data set have been considered for different processes with increasing, decreasing and random order of burst time respectively.

### 3.4 Performance Metrics

The significance of our performance metrics for experimental analysis is as follows:

1) *Turnaround time (TAT)*: For the better performance of the algorithm, average turnaround time should be less.
2) *Waiting time (WT)*: For the better performance of the algorithm, average waiting time should be less.
3) *Number of Context Switches (CS):* For the better performance of the algorithm, the number of context switches should be less.

### 3.5 Results Obtained
### Case 1:  Increasing Order of Burst Time

We assume five processes arriving at time=0, with increasing burst time (P1=5,P2=12,P3=16,P4=21,P5=23) and priority (P1=2, P2=3, P3=1, P4=4, P5=5). Table-3.1, Table-3.2,Table-3.3 show the output using algorithm in paper[6],[8] and our new proposed algorithm respectively.  Table3.4 shows the comparison between table 3.1, 3.2 & 3.3.

**Table 3.1 for Data in Increasing Order as Per Paper [6]**

| PROCESS ID | BURST TIME | PRIORITY | OTS | PC | SC | CSC | ITS |
|---|---|---|---|---|---|---|---|
| P1 | 5 | 2 | 4 | 0 | 0 | 1 | 5 |
| P2 | 12 | 3 | 4 | 0 | 0 | 0 | 4 |
| P3 | 16 | 1 | 4 | 1 | 0 | 0 | 5 |
| P4 | 21 | 4 | 4 | 0 | 0 | 0 | 4 |
| P5 | 23 | 5 | 4 | 0 | 0 | 0 | 4 |

| P1 | P2 | P3 | P4 | P5 | P2 | P3 | P4 | P5 |
|----|----|----|----|----|----|----|----|----|

0    5    9    14    18    22    26    31    35

| P 5 | P2 | P3 | P4 | P5 | P3 | P4 | P5 | P4 |
|-----|----|----|----|----|----|----|----|----|

39    43    48    52    56    57    61    65

| P4 | P5 | P4 | P5 | ... |
|----|----|----|----|-----|

69    73    74    77

**Fig 3.1: Gantt Chart For Table3.1**

**Table 3.2 For Data In Increasing Order As Per Paper [8]**

| PROCESS ID | SC | ITS | ROUNDS | | | | |
|------------|----|----|-----|-----|-----|-----|-----|
| | | | 1ST | 2ND | 3RD | 4TH | 5TH |
| P1 | 0 | 5 | 5 | 0 | 0 | 0 | 0 |
| P2 | 0 | 4 | 2 | 3 | 7 | 0 | 0 |
| P3 | 0 | 5 | 3 | 5 | 8 | 0 | 0 |
| P4 | 0 | 4 | 2 | 3 | 5 | 8 | 3 |
| P5 | 0 | 4 | 2 | 3 | 5 | 8 | 5 |

| P1 | P2 | P3 | P4 | P5 | P3 | P2 | P4 | P5 |
|----|----|----|----|----|----|----|----|----|

0    5    7    10    12    14    17    22    25

| P5 | P2 | P3 | P4 | P5 | P4 | P5 | P4 | P5 |
|----|----|----|----|----|----|----|----|----|

28    35    43    48    53    61    69
72    77

**Fig 3.2 Gantt Chart For Table 3.2**

**Table 3.3 For Data In Increasing Order As Per Our Proposed Algorithm**

| PROCESS ID | OTS | CSC | ITS | ROUNDS | | | | |
|------------|-----|-----|-----|-----|-----|-----|-----|-----|
| | | | | 1ST | 2ND | 3RD | 4TH | 5TH |
| P1 | 7 | 5 | 12 | 5 | 0 | 0 | 0 | 0 |
| P2 | 5 | 0 | 5 | 3 | 5 | 4 | 0 | 0 |
| P3 | 14 | 2 | 17 | 9 | 7 | 0 | 0 | 0 |
| P4 | 4 | 0 | 4 | 2 | 3 | 5 | 8 | 3 |
| P5 | 3 | 0 | 3 | 2 | 3 | 5 | 8 | 5 |

| P1 | P2 | P3 | P4 | P5 | P2 | P3 | P4 | P5 |
|----|----|----|----|----|----|----|----|----|

0    5    8    17    19    21    28    33    36

| P5 | P3 | P4 | P5 | P4 | P5 | P4 | P5 |
|----|----|----|----|----|----|----|----|

39    43    48    53    61    69    72    77

**Fig 3.3: Gantt Chart For Table 3.3**

**Table 3.4 Comparison Among The Algorithm In   Paper [6], [8] And Proposed Method**

| Algorithm | Avg TAT | Avg WT | CS |
|-----------|---------|--------|-----|
| In paper[6] | 51.2 | 35.8 | 19 |
| In paper [8] | 46.4 | 31 | 17 |
| In Proposed method | 46 | 30.6 | 15 |

## CASE 2: Decreasing Order of Burst Time

We assume five processes arriving at time=0, with decreasing burst time  (P1=31,P2=23,P3=16,P4=9,P5=1)  and  priority (P1=2,  P2=1,  P3=4,  P4=5,  P5=3). Table-3.5, Table-3.6, Table-3.7 show the output using algorithm in paper [6], paper [8] and the new proposed algorithm. Table 3.8 shows the comparison between tables 3.5, 3.6 & 3.7.

**Table 3.5 for Data in Decreasing Order as Per Paper [6]**

| PROCESS ID | BURST TIME | PRIORITY | OTS | PC | SC | CSC | ITS |
|---|---|---|---|---|---|---|---|
| P1 | 31 | 2 | 4 | 0 | 0 | 0 | 4 |
| P2 | 23 | 1 | 4 | 1 | 1 | 0 | 6 |
| P3 | 16 | 4 | 4 | 0 | 1 | 0 | 5 |
| P4 | 9 | 5 | 4 | 0 | 1 | 0 | 5 |
| P5 | 1 | 3 | 4 | 0 | 1 | 0 | 1 |

| P1 | P2 | P3 | P4 | P5 | P1 | P2 | P3 | P4 |
|---|---|---|---|---|---|---|---|---|

0    4    10    15    20    21    25    31    36

| P4 | P1 | P2 | P3 | P1 | P2 | P1 | P1 | P1 |
|---|---|---|---|---|---|---|---|---|

40    44    50    56    60    65    69    73

| P1 | P1 |
|---|---|

77    80

**Fig 3.4: Gantt Chart For Table 3.5**

**Table 3.6 For Data In Decreasing Order As Per Paper [8]**

| PROCESS ID | SC | ITS | ROUNDS | | | |
|---|---|---|---|---|---|---|
| | | | 1ST | 2ND | 3RD | 4TH |
| P1 | 0 | 4 | 2 | 3 | 5 | 21 |
| P2 | 1 | 6 | 6 | 12 | 5 | 0 |
| P3 | 1 | 5 | 5 | 11 | 0 | 0 |
| P4 | 1 | 5 | 5 | 4 | 0 | 0 |
| P5 | 1 | 1 | 1 | 0 | 0 | 0 |

| P1 | P2 | P3 | P4 | P5 | P1 | P2 | P3 | P4 |
|---|---|---|---|---|---|---|---|---|

0   2    8    13    18    19    22    34    45

| P4 | P1 | P2 | P1 |
|---|---|---|---|

49    54    59    80

**Fig 3.5: Gantt chart for Table 3.6**

**Table 3.7 for Data in Decreasing Order as Per Proposed Algorithm**

| PROCESS ID | OTS | CSC | ITS | ROUNDS | | | |
|---|---|---|---|---|---|---|---|
| | | | | 1ST | 2ND | 3RD | 4TH |
| P1 | 8 | 0 | 8 | 4 | 6 | 9 | 12 |
| P2 | 16 | 5 | 23 | 23 | 0 | 0 | 0 |
| P3 | 4 | 0 | 5 | 5 | 11 | 0 | 0 |
| P4 | 3 | 0 | 4 | 4 | 5 | 0 | 0 |
| P5 | 5 | 0 | 1 | 1 | 0 | 0 | 0 |

| P5 | P4 | P3 | P2 | P1 | P4 | P3 | P1 | P1 | P1 |
|---|---|---|---|---|---|---|---|---|---|

0    1    5    10    33    37    42    53    59    68   80

**Fig3.6: Gantt Chart For Table 3.7**

**Table 3.8 Comparison Among The Algorithm In Paper [6], [8] And Proposed Method**

| Algorithm | AVG TAT | Avg WT | CS |
|---|---|---|---|
| In paper[6] | 54 | 38 | 12 |
| In paper[8] | 50.4 | 34.4 | 12 |
| In proposed method | 41.8 | 25.8 | 7 |

## CASE 3: Random Order of Burst Time

We assume five processes arriving at time=0, with random burst time (P1=11,P2=53,P3=8,P4=41,P5=20) and priority (P1=3, P2=1, P3=2, P4=4, P5=5). Table 3.9, Table-3.10, Table-3.11 show the output using algorithm in paper [6], paper [8] and our new proposed algorithm. Table 3.12 shows the comparison between tables 3.9, 3.10 and 3.11 respectively.

**Table 3.9 For Data In random Order as Per Paper [6]**

| PROCESS ID | BURST TIME | PRIORITY | OTS | PC | SC | CSC | ITS |
|---|---|---|---|---|---|---|---|
| P1 | `11 | 3 | 4 | 0 | 0 | 0 | 4 |
| P2 | 53 | 1 | 4 | 1 | 0 | 0 | 5 |
| P3 | 8 | 2 | 4 | 0 | 1 | 3 | 8 |
| P4 | 41 | 4 | 4 | 0 | 0 | 0 | 4 |
| P5 | 20 | 5 | 4 | 0 | 1 | 0 | 5 |

| P1 | P2 | P3 | P4 | P5 | P1 | P2 | P4 | P5 | P1 |
|---|---|---|---|---|---|---|---|---|---|

0    4    9    17    21  26    30  35    39  44

| P1 | P2 | P4 | P5 | P2 | P4 | P5 | P2 | P4 | P2 |
|---|---|---|---|---|---|---|---|---|---|

47    52    56    61    66  70    75    80  84    89

| P4 | P2 | P4 | P2 | P4 | P2 | P4 | P2 | P4 | P2 |
|---|---|---|---|---|---|---|---|---|---|

93    98    102   107  111  116   120  125  130  133

**Fig3.7: Gantt Chart For Table 3.9**

**Table 3.10 For Data In Random Order As Per Paper [8]**

| PROCESS ID | SC | ITS | ROUNDS | | | | | |
|---|---|---|---|---|---|---|---|---|
| | | | 1ST | 2ND | 3RD | 4TH | 5TH | 6TH |
| P1 | 0 | 4 | 2 | 3 | 6 | 0 | 0 | 0 |
| P2 | 0 | 5 | 3 | 5 | 8 | 12 | 18 | 7 |
| P3 | 1 | 8 | 8 | 0 | 0 | 0 | 0 | 0 |
| P4 | 0 | 4 | 2 | 3 | 5 | 8 | 12 | 11 |
| P5 | 1 | 5 | 5 | 10 | 5 | 0 | 0 | 0 |

| P1 | P2 | P3 | P4 | P5 | P1 | P2 | P4 | P5 | P1 |
|---|---|---|---|---|---|---|---|---|---|

0    2    5    13    15  20    23  28  31    41

| P1 | P2 | P4 | P5 | P2 | P4 | P2 | P4 | P2 | P4 |
|---|---|---|---|---|---|---|---|---|---|

47    55    60    65    77    85    103  115 122 133

**Fig 3.8: Gantt Chart For Table 3.10**

**Table 3.11 For Data In Random Order As Per Proposed Algorithm**

| PROCESS ID | OTS | CSC | ITS | ROUNDS | | | | |
|---|---|---|---|---|---|---|---|---|
| | | | | 1ST | 2ND | 3RD | 4TH | 5TH |
| P1 | 10 | 1 | 11 | 6 | 5 | 0 | 0 | 0 |
| P2 | 31 | 21 | 53 | 27 | 26 | 0 | 0 | 0 |
| P3 | 16 | 8 | 25 | 8 | 0 | 0 | 0 | 0 |
| P4 | 8 | 0 | 8 | 4 | 6 | 9 | 15 | 7 |
| P5 | 6 | 0 | 7 | 7 | 13 | 0 | 0 | 0 |

| P3 | P1 | P5 | P4 | P2 | P1 | P5 | P2 | P4 | P4 |
|---|---|---|---|---|---|---|---|---|---|

0    8    14    21    25    52    57    70    96    102

| P4 | P4 | P4 |
|---|---|---|

111 126 133

**Fig 3.9: Gantt Chart For Table 3.11**

**Table 3.12 Comparison Among The Algorithms In Paper [6], [8] And   Proposed Method**

| Algorithm | Avg TAT | Avg WT | CS |
|---|---|---|---|
| In Paper[6] | 80.8 | 54.2 | 29 |
| In  Paper[8] | 76 | 49.2 | 18 |
| In Proposed Method | 72.8 | 36.2 | 9 |

**Fig.3.10:**Comparison among the average turn around time .average waiting time and number of context switches of algorithms in[6],[8] and proposed method for data in increasing order of burst time

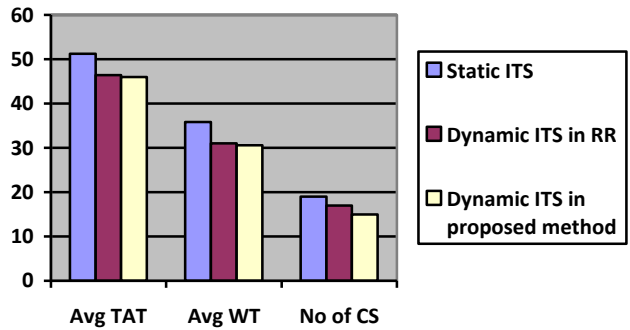

**Fig-3.11:**Comparison among the average turn around time .average waiting time and number of context switches of algorithms in[6],[8] and proposed method for data in decreasing order of burst time.

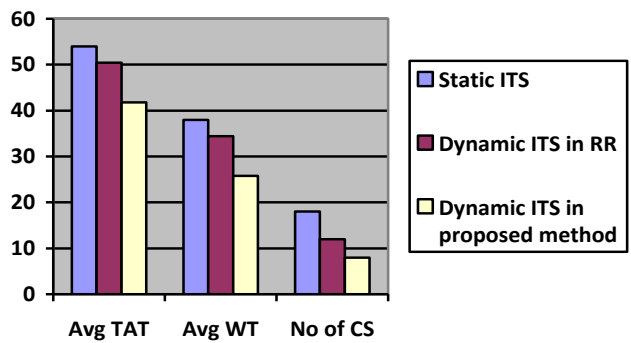

**Fig-3.12** Comparison among the average turn around time .average waiting time and number of context switches of algorithms in[6],[8] and proposed method for   data in random order of burst time.

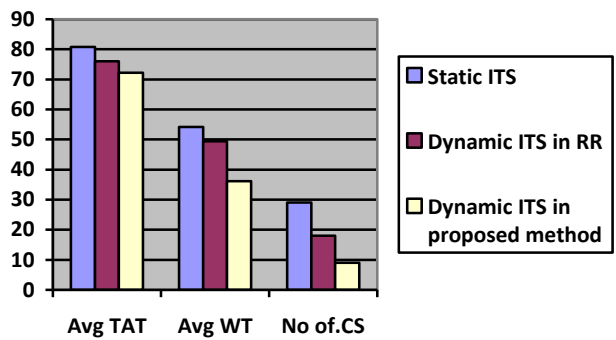

# 4. CONCLUSION AND FUTURE WORK

It is concluded from the above experiments that the proposed algorithm performs better than the algorithm proposed by C.Yaashuwanth et.al [6] and prof R. Mohanty and Prof H.S.Behera et.al [8]  in terms of performance metrics such as average waiting time,average turn around time and total number of context switches and the time and space complexity is reduced.

Future work can be enhanced to implement the proposed algorithm for adaptive and hard real time systems.